# MEASUREMENT AND CHARACTERIZATION OF THE STATIONARY NOISE IN NARROWBAND POWER LINE COMMUNICATION

Raja Alaya[1] and Rabah Attia[2]

[1] Tunisian Polytechnic School, University of Carthage, Tunisia
raja.alaya@ept.rnu.tn, Rabah.attia@ept.rnu.tn

## ABSTRACT

*Understanding the interference scenario in power lines network is a key step to characterize the power line communication (PLC) system. This paper focuses on the characterization and modelling of the stationary noise in Narrowband PLC. Measurement and analysis of noise is carried out in the Tunisian outdoor Low Voltage (LV) power line network in the frequency band below 500 kHz. Based on existing models and measurements results, a parametric model of noise is proposed; the model parameters are statistically studied.*

## 1. INTRODUCTION

In the last few years, narrowband (i.e., frequency range between 9 and 500 kHz) power line communication (NB-PLC) technology has attracted the industry and tends to be a promising way of information exchange. This technology is widely used in Smart Grid (SG) applications as a communication infrastructure for advanced metering infrastructures communications, automatic meter reading and demand response [1]. The main advantage of PLC is to use the electrical network for the communication which is an existing infrastructure and it potentially covers the entire country under study. This permits to avoid additional wiring and have a quick deployment. However the existing power lines were originally not designed for signal transmission, but only for electricity delivery for end customer. Otherwise LV power lines present a very harsh communications media which suffers from several kinds of disturbances, such as the stationary noise and the impulsive noise. The presence of the noise in PLC system is a crucial factor that affects the transmitted signals. In fact, the performance of a communication system is directly related to the noise level [2]. To this extent, it is useful to have as much knowledge as possible of the noise by identifying the sources of interference and developing a mathematical model which adequately describes noise characteristics in power line communication networks.

To characterize noise and develop its models, statistical knowledge of noise parameters are required that can acutely describe noise present in power line, which can be obtained from experimental measurement. Therefore a lot of measurement campaigns should be done.

A lot of works have already characterized the impulsive noise and stationary noise. However they are primarily proposed in the case of Indoor at the Broadband frequency [3, 4, 5]. The available noise measurements carried out in LV distribution networks and Narrowband frequency are mostly conducted to only describe the impulsive noise, and there has not been much attention to the stationary noise. Even if many studies are available in the literature [6, 7, 8], there is no works have been performed on the Tunisia distribution LV network. In this context, the manuscript proposes the following two main contributions: the first one is measurement and analysis of the stationary noise present in the LV distribution networks in Tunisia. The second contribution is to propose a statistical parametric model based on the analysis of the measured noise. The parameters of the model are considered as random variables approximated by their corresponding statistical distributions. This model can be used for advanced studies as a noise generator.

In this manuscript, the state of art is described in Section 2. In particular, the sources of interferences in PLC and the existing models of noise are presented. Section 3 explains the experimental measurements' setup and the main results: the spectral analysis of the NB-PLC noise in the various locations. Section 4 presents characterization and modelling of the noise in the frequency domain. Finally, the paper is concluded in Section 5.

## 2. State of art

### 2.1. Sources of interferences

Unlike the other wired transmission structures, noise present in power-line cannot be described with an Additive White Gaussian Noise (AWGN), and it is hard to characterize it with one universal model. In fact, the PLC channel is as a hostile environment for data transmission and the noise scenario is rather complicated and exhibit quite different behaviours due to the presence of different sources of disturbance. As can be seen in Figure 1, power line noise can be separated in the following classes according to their spectral and time behaviours [9], [10]:

1) Colored background noise: it is always present in the network. It represents the summation of the noise generated by different noise sources. Its Power Spectral Density (PSD) increases towards lower frequencies.
2) Narrow-band noise: it is a radio frequency interference primary originates from the broadcasting stations. The amplitude can be changed in dependence on time and place.
3) Asynchronous periodic impulsive noise. This kind of noise can be considered as cyclostationary and synchronized with the mains; it is mostly caused by switched mode power supplies.
4) Synchronous periodic impulsive noise. It is a cyclostationary noise that synchronous with the mains. It is mainly caused by switching actions of rectifier diodes which occurs synchronously with the mains cycle.
5) Asynchronous impulsive noise. It is mainly generated by switching transients in the network; it is the most harmful noise source.

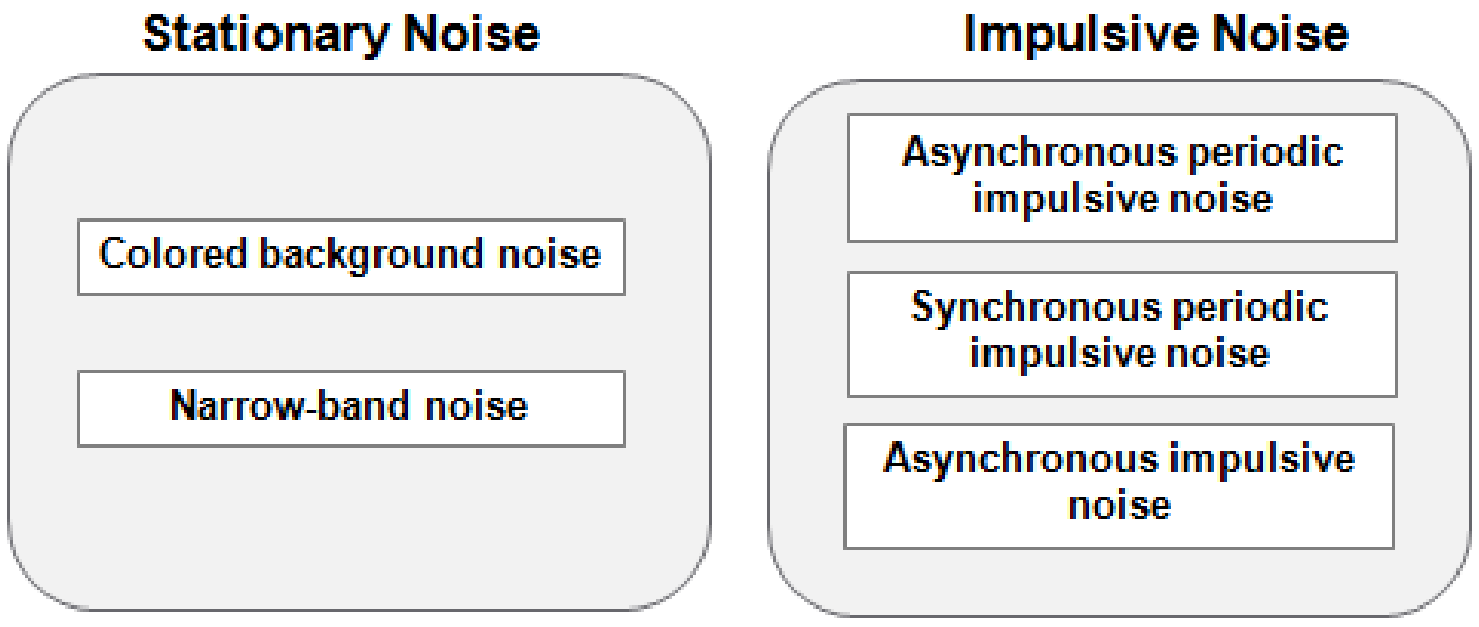

Figure 1. Classification of Noise on PLC Channels.

According to the literature [11], the amplitudes of the two first types vary slowly over time from seconds to even hours, so that they can be summarized as stationary noise. Then again, the amplitudes of the last three types change rapidly; they are defined as a set of single pulses or bursts with short durations from microseconds to milliseconds. Therefore, they can be regarded as impulsive noise. In this paper, our interest is mainly focused in types (1) and (2), while types (3), (4) and (5) will be studied in future works.

### 2.2. Noise Models

Noise models available in the technical literature are obtained based on empirical measurements in the frequency or time domain. Generally, background noise is modelled in the frequency domain [12], while the impulsive noise is modelled in both frequency and time domains [13]. The well existing techniques to model the noise are developed by fitting the measured noise PSD into certain functions of frequency. Moreover, noise models are obtained based on fitting the statistical properties of power line noise. Many researchers have already studied noise of type (1) and type (2) and many noise models are available in the literature, the most famous and well accepted PLC noise models for the colored background noise are the Esmalian model [14]and the OMEGA model [15]. The OMEGA model is a frequency domain model for the colored background noise. This model, based on PSD fitting noise, is characterized by a logarithmic decay function. The Esmalian model is also a frequency domain model for the colored background noise. It is based on measurements and depicts the background noise as a power frequency dependent function. Another model proposed by Philipps [16], suppose that the noise PSD is a first order exponential function, where the noise PSD decreases as the frequency gets larger. For the Narrowband noise, the most common model is a Gaussian function represented by summation of various signals. In [17] the Narrowband noise is modelled as the sum of multiple sine signals with different amplitudes at specific frequencies. The stationary noise model is basically regarded as the superposition of background noise model and narrowband noise model [18].

## 3. NOISE MEASUREMENT SETUP AND RESULTS

### 3.1. Experimental measurement setup

The measurement setup of the frequency domain noise is performed with a spectrum analyzer GSP-830 from GW INSTEK, configured with a resolution bandwidth of 3 kHz, and a sweeping time of 80 ms on the total band up to 500 kHz. It is important to note that all measurements are carried out in the same configuration. As shown in Figure 2, the spectrum analyzer is connected to the power line through a coupling unit which is designed by the authors in [19]. The coupling unit, as the name implies couple and decouples the high frequency signal from or to the power line. In other words, it blocks the 50/60 Hz current from entering the measurement instrument, it also prevents the high voltage of the mains from damaging the measurement instrument.

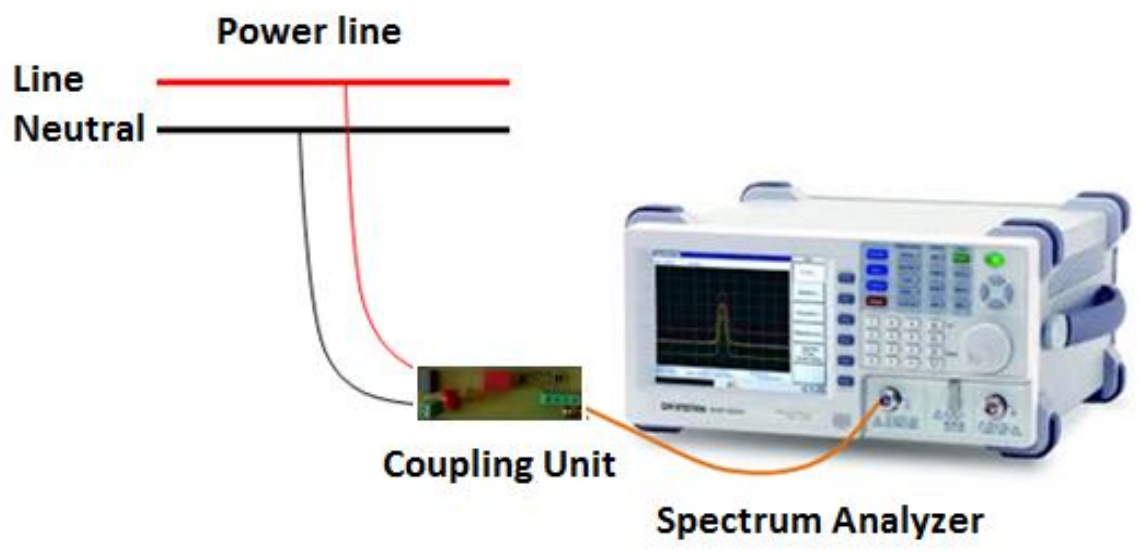

Figure 2. Experimental setup to measure the stationary noise

The schematic diagram of the coupling unit is shown in Figure 3.

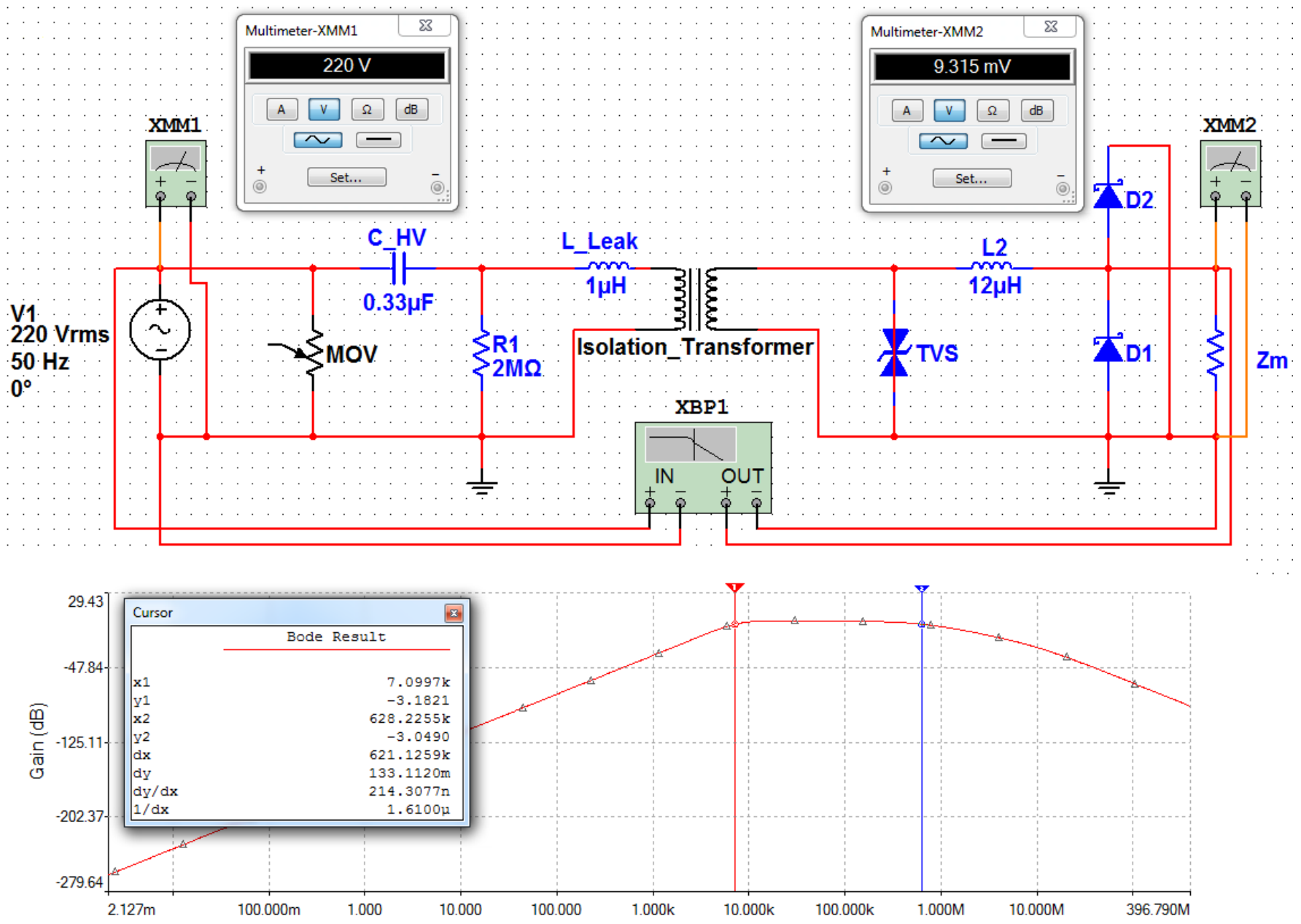

Figure 3. Designed Coupling Unit and Simulated band-pass filter.

During two weeks, many measurements were performed at diverse times of the day, in three typical urban sites in Tunisia: the first one, designed in the following site (S1), is an underground network; this site is modern and simple. The second one is dense, designed in the following site (S2), it is a mix of both underground and overhead lines; the customers in this site can be commercial or residential. The third one, designed in the following site (S3), is extremely dense and complex; it is an overhead network in which all the customers are residential.

## 3.2 Experimental Results and Discussion

Figure 4 and Figure 5 illustrates respectively the evolution of the noise PSD in the three typical sites in the customer side and the evolution of the noise PSD in the transformer substation.

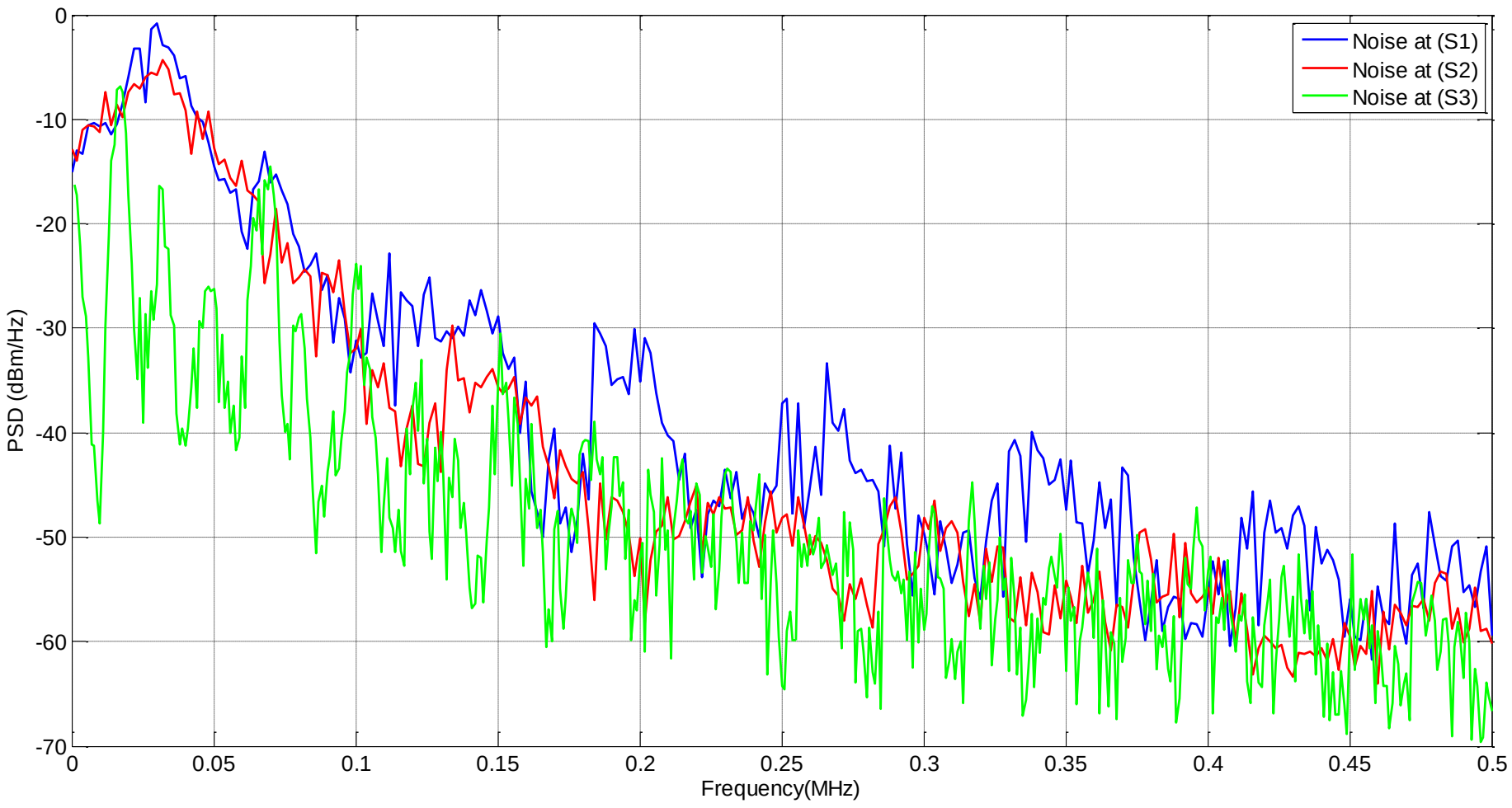

Figure 4. Noise PSD at the customer side at the meter in the three typical sites.

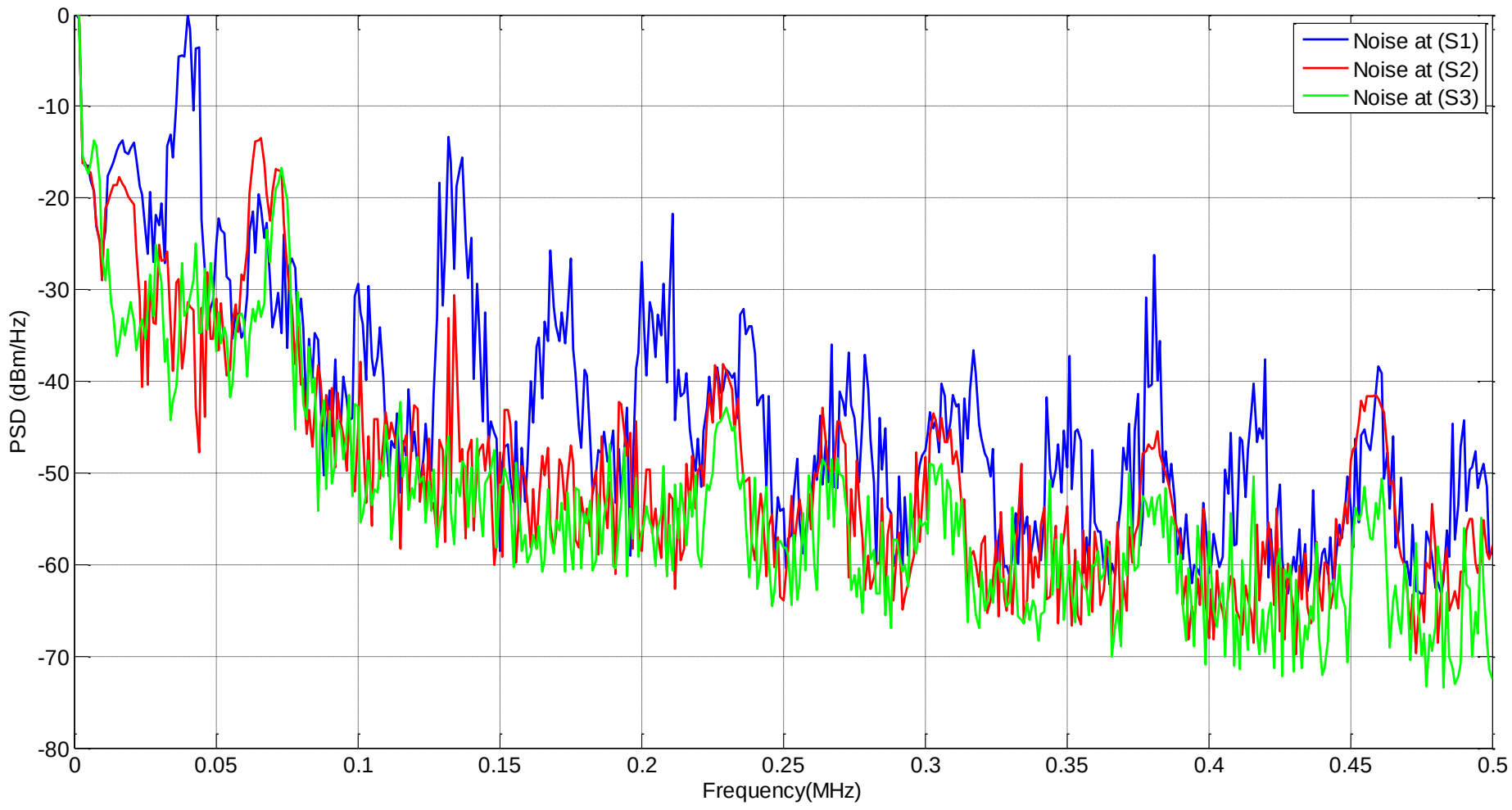

Figure 5. Noise power spectral density (PSD) in the transformer substation in the three typical sites.

It is noted that the noise has a frequency dependent nature. The noise level is higher at the lower frequencies and the level decreases as the frequency increases. The shape of the noise is a decreasing function of frequency with the presence of narrowband interferences. The interferences appear in different frequencies with random level and bandwidth. This is caused by the presence of many sources of low-frequency noise in the power network.

Comparing the noise level in the three sites, we can say that the noise level measured at the site (S3) is significantly higher than the noise level measured at the site (S1) and the site (S2), this may be explained by the important number of loads connected at the same time and topology of the network which is very complex and extremely dense. Comparing the noise measured at the site (S1) and at the site (S2), we can say that the number of interferences increases at the site (S2), because the usage of electrical appliances is significant, since there is a commercial consumer in this site. However, it can be noted that the shapes of the measured noises are in general similar.

Finally, in comparison with the customer side, the number and the depth of the interferences at the transformer side are high, because the number of loads (customers) connected to the transformer substation is more important. Hence, noise measured in the transformer substation is the sum of noise generated by all domestics loads connected to the same branch.

## 4. Noise characterization and modelling

The stationary noise in the PLC environment is regarded as the superposition of the background noise and the narrowband interferences, which is the basis of the following modelling.

### 4.1. Colored Background Noise

Based on empirical measurements, the noise model is obtained by fitting the measured noise PSD into certain functions of frequency. In the literature, several models of colored background noise are proposed [14, 15].

Since, the background noise PSD is a decreasing function of frequency, the author use the Esmalian model, where the noise is considered Gaussian and it is described as follow:

$$N_{CB} = af^{b} + c \qquad (1)$$

Where $N_{CB}$ stands for the Colored Background noise PSD in dBm/Hz, $f$ denotes the frequency in Hz, a, b and c are the model parameters derived from measurements. The parameter a controls the noise floor, b controls the value of noise PSD at starting frequency and c controls the form of the frequency dependent decay.

Figure 6 illustrates the corresponding fitting (bold curve) of the measured background noise in two cases.

It is observed that for stronger noise, we have the worst case with parameters values: a = -66.76, b = 0.3942, c = -12.9. For weaker noise, we have the best case with values: a = -76.95, b = 0.25, c = 0.03.

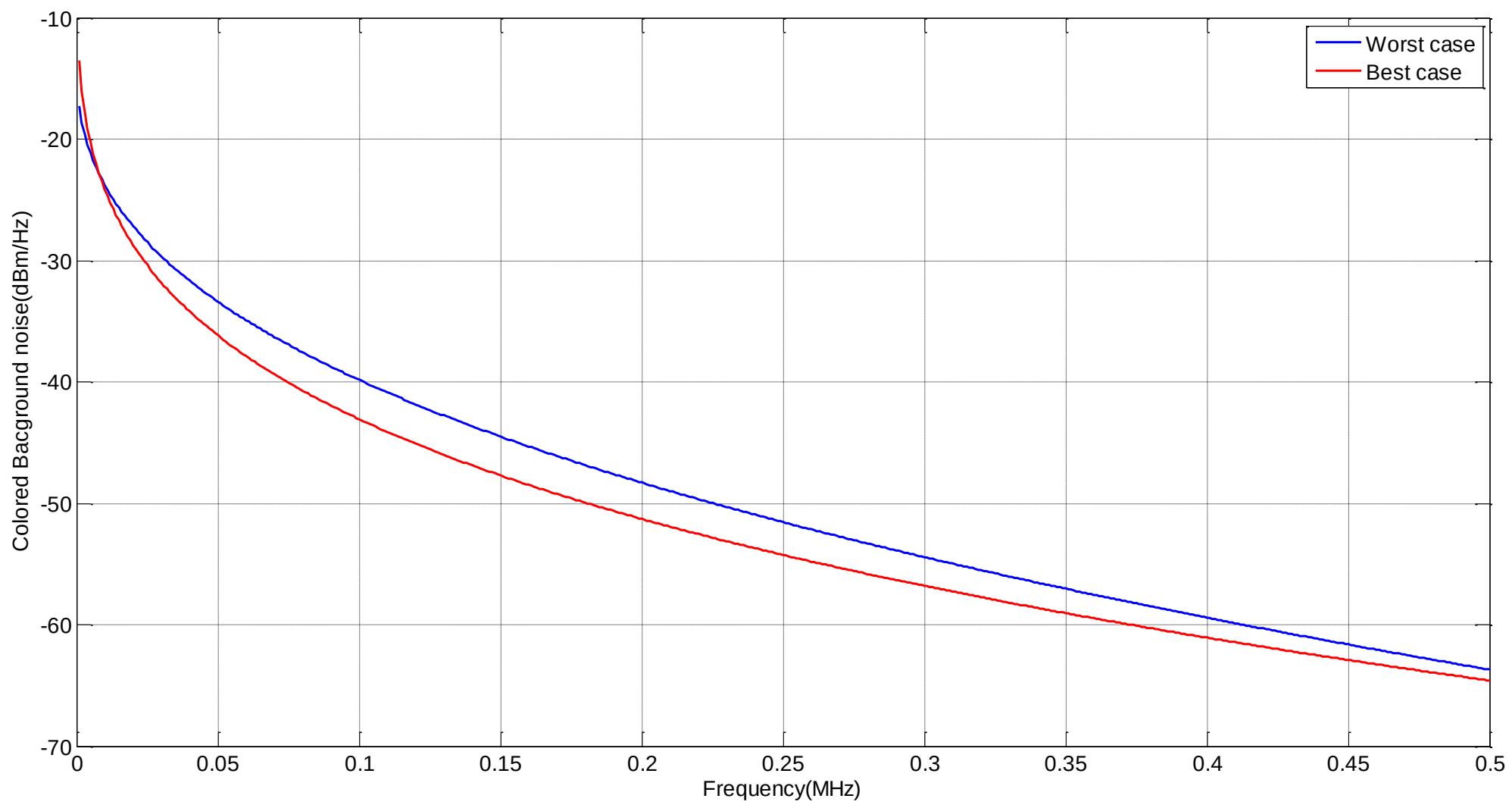

Figure 6. Colored background noise model.

## 4.2. Narrowband noise

The narrowband noise is the sum of different narrowband noises. Therefore the proposed model is a summation of a parametric Gaussian function expressed by:

$$N_{NI} = \sum_{i=1}^{N} A_i \exp\left(-\frac{(f - f_i)^2}{2\sigma_i^2}\right) \qquad (2)$$

Where $N_{NI}$ is the number of narrowband interferences, $A_i$ is the amplitude, $f_i$ for the center frequency of each interference and $\sigma_i$ represents the Gaussian function standard deviation which controls the bandwidth of the narrowband interference.

Figure 7 shows the resulting narrowband noise.

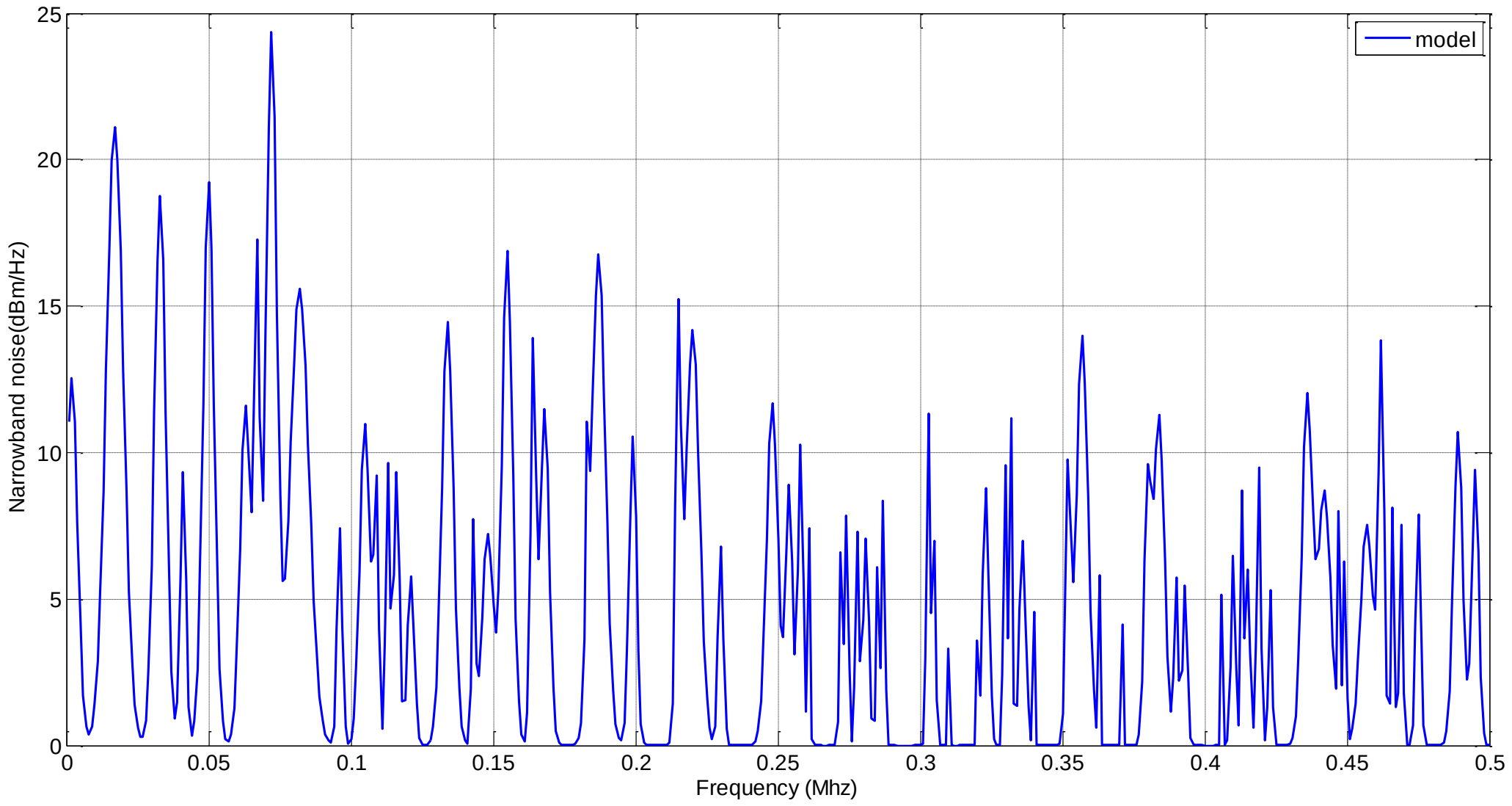

Figure 7. Simulated narrowband noise model.

The parameters of the Gaussian function are derived from measurements by doing the difference between the measured stationary noise PSD and the model of the colored background noise PSD expressed in (1).

When designing a noise model, it is crucial to acquaint the statistical behaviour of the parameters $A_i$, $f_i$ and $\sigma_i$.

Figure 8 illustrates the Cumulative distribution function (CDF) of the narrowband interferences amplitudes $A_i$; it is observed that this parameter follows Gamma distribution.

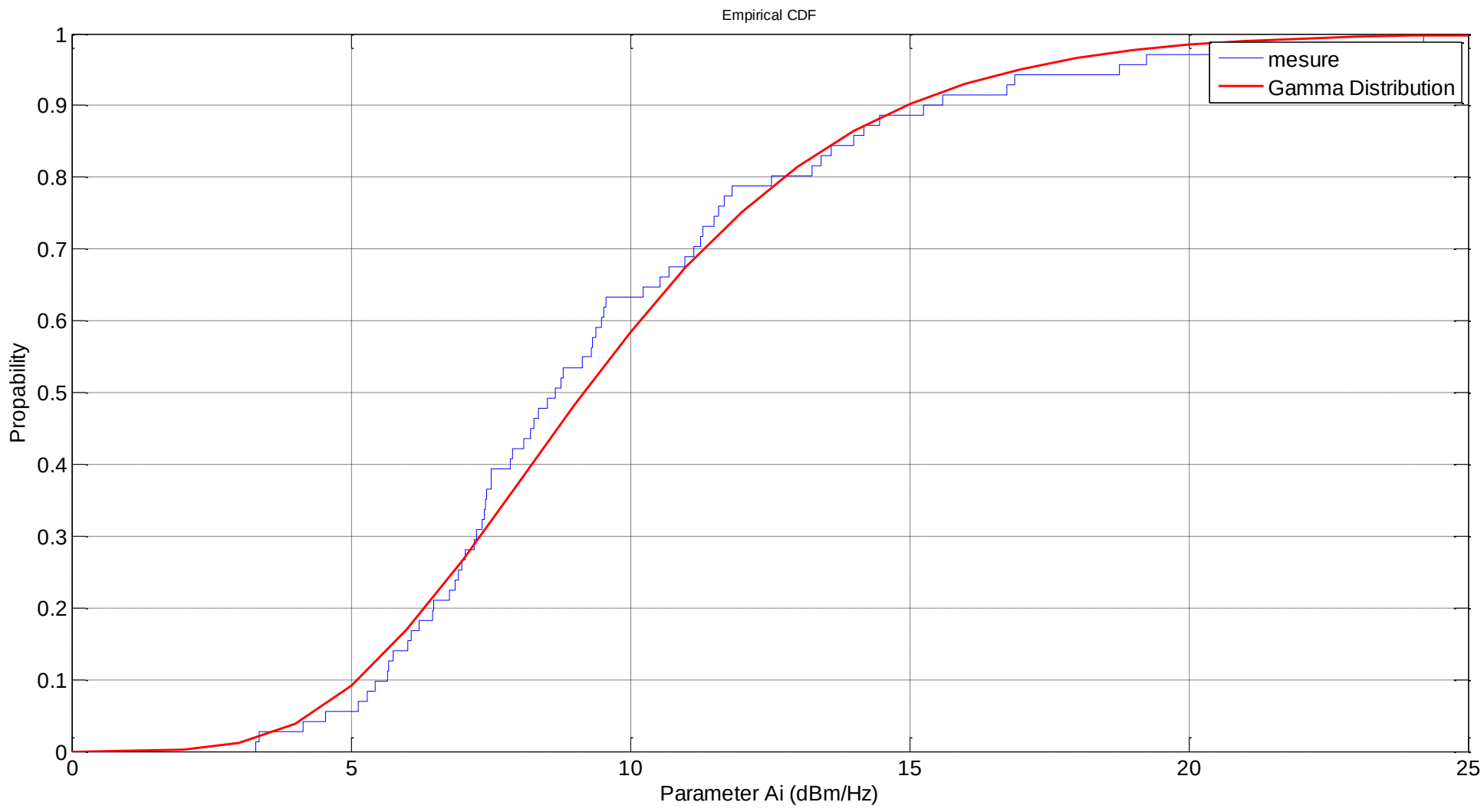

Figure 8. CDF of the parameter $A_i$

Figure 9, shows the CDF of the narrowband interferences center frequency $f_i$ which is normally distributed.

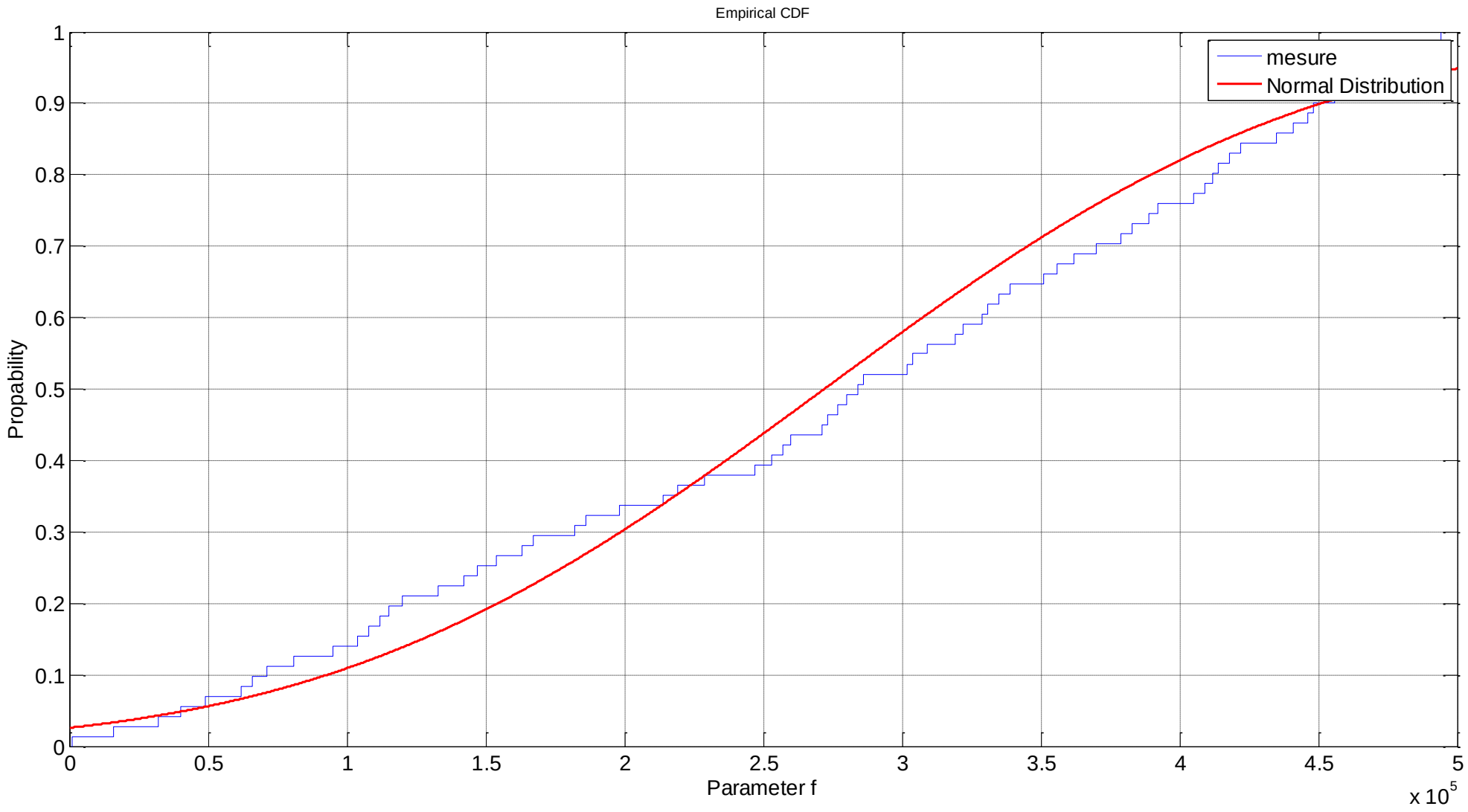

Figure 9. CDF of the parameter $f_i$

Figure 10 represents the CDF of the parameter $\sigma_i$ which control the narrowband interferences bandwidth. It is noted that $\sigma_i$ follows a Gamma distribution.

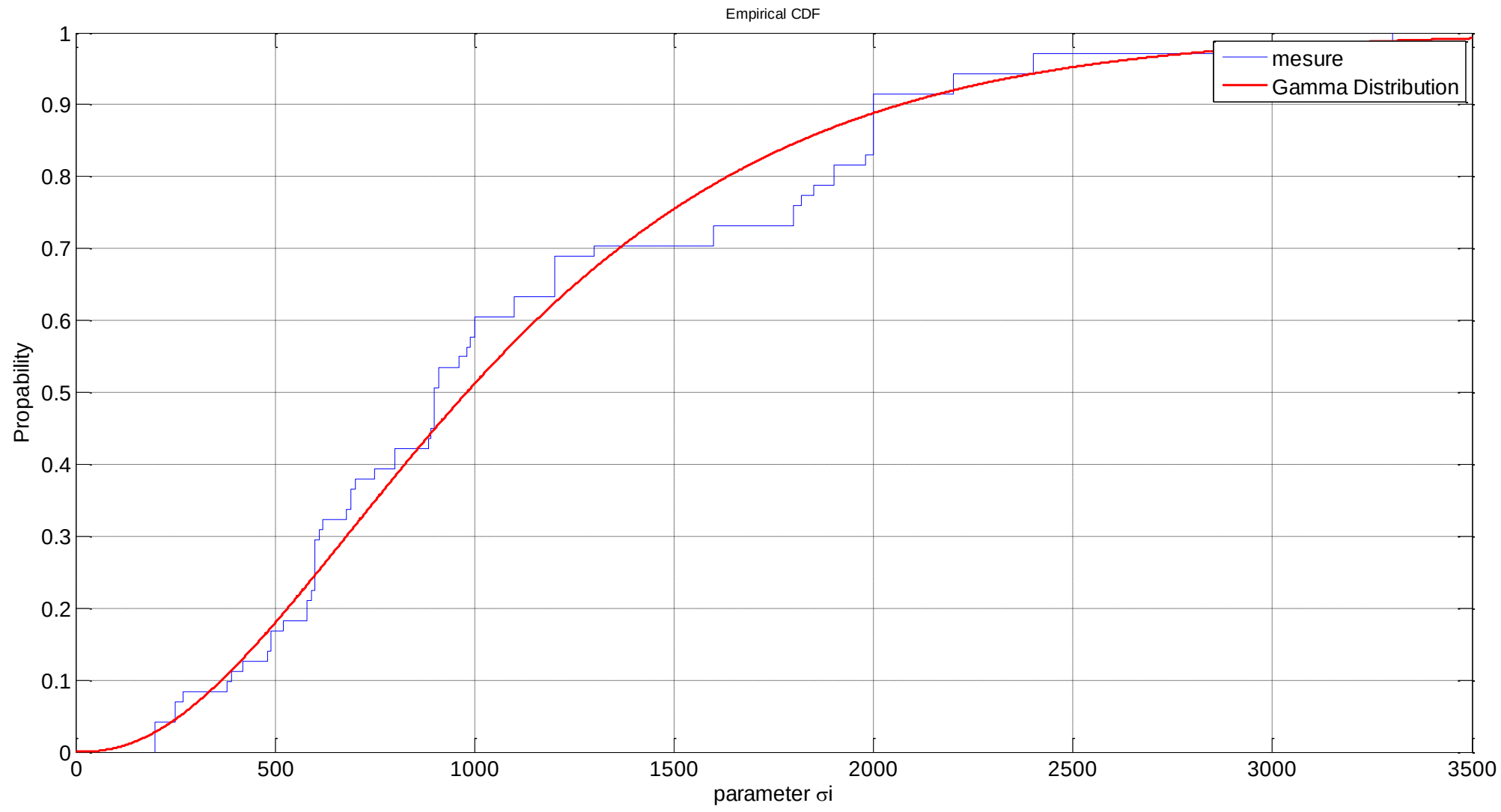

Figure 10. CDF of the parameter

From the statistical analysis of the parameters $A_i$ , $f_i$ and $\sigma_i$, we can notice that the narrowband noise model parameters are not unambiguous defined and they have statistical properties. The CDF of the narrowband interferences amplitude $A_i$ follows a Gamma distribution with shape parameter equal to 6.09 and scale parameter equal to 1.58. The parameter $f_i$ is well fitted by a normal distribution with a mean value equal to 2.72 and a standard deviation equal 1.39. The parameter $\sigma_i$ is also approximated by Gamma distribution with shape parameter equal to 2.54 and scale parameter equal to 440.99.

### 4.3. Stationary noise

As it is depicted above, the stationary noise is considered as the superposition of the colored background noise and the narrowband noise because they remain stationary over time. The PSD of the stationary noise is expressed by:

$$N_S = N_{CB} + N_{NI} \qquad (3)$$

Where $N_S$ is the stationary noise, $N_{CB}$ is the colored background noise and $N_{NI}$ is the narrowband noise.

Figure 11 illustrates the PSDs of an example of the measured noise and the proposed model.

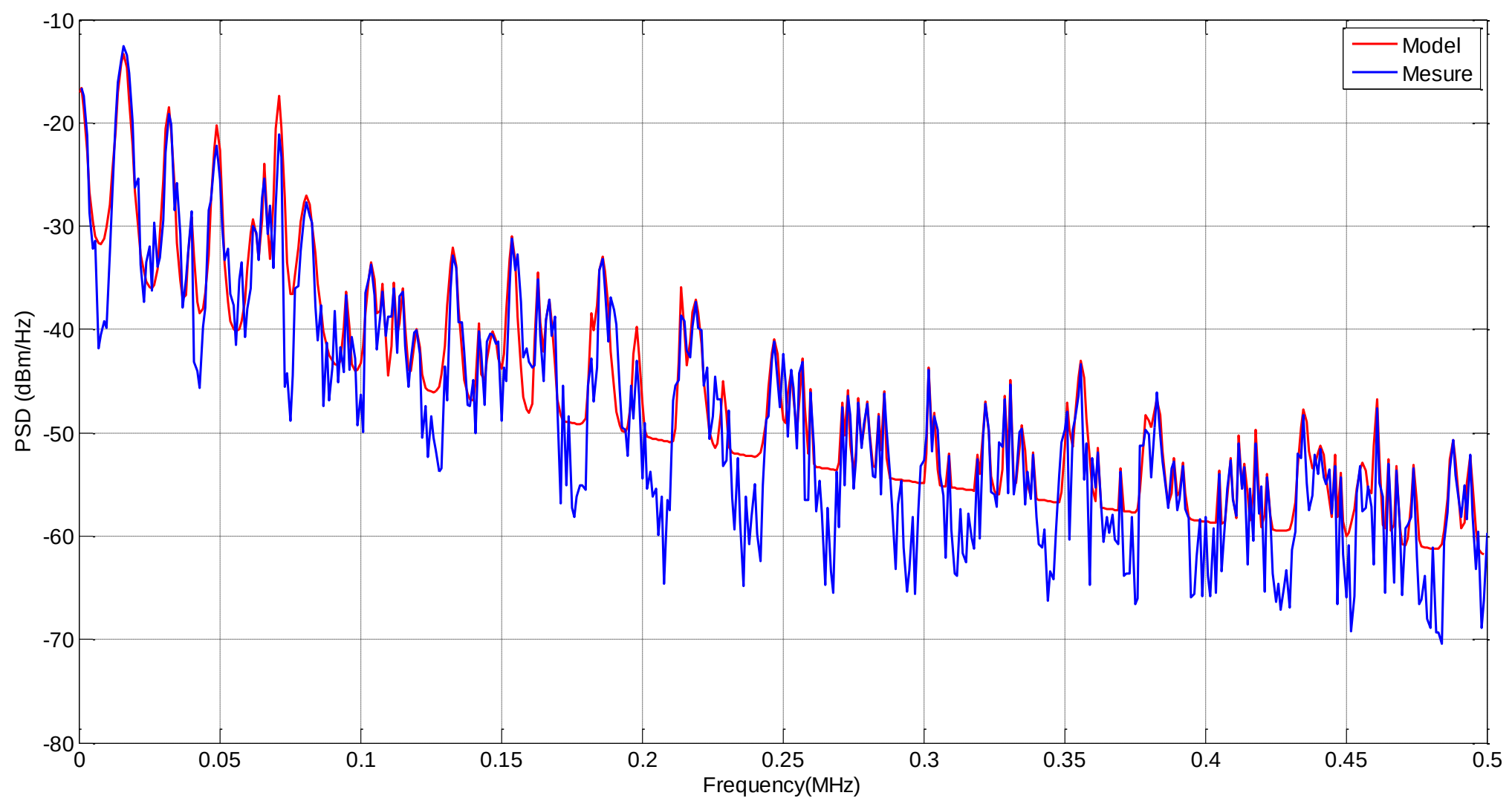

Figure 11. PSD of the Stationary modelled and measured noise.

It is shown that the proposed model is in good agreement with the measured PSD. The red curve shows the stationary noise model following expression (3) and corresponding to the measurement example. We can see the decreasing trend of the background noise with increasing frequencies and the presence of narrowband interferences with higher levels in the very low frequency range.

In order to validate the proposed model we statistically process the simulated noise and compare results with measured noise.

Figure 12 and Figure 13 depict the PDFs and CDFs of the modelled and measured noise amplitude.

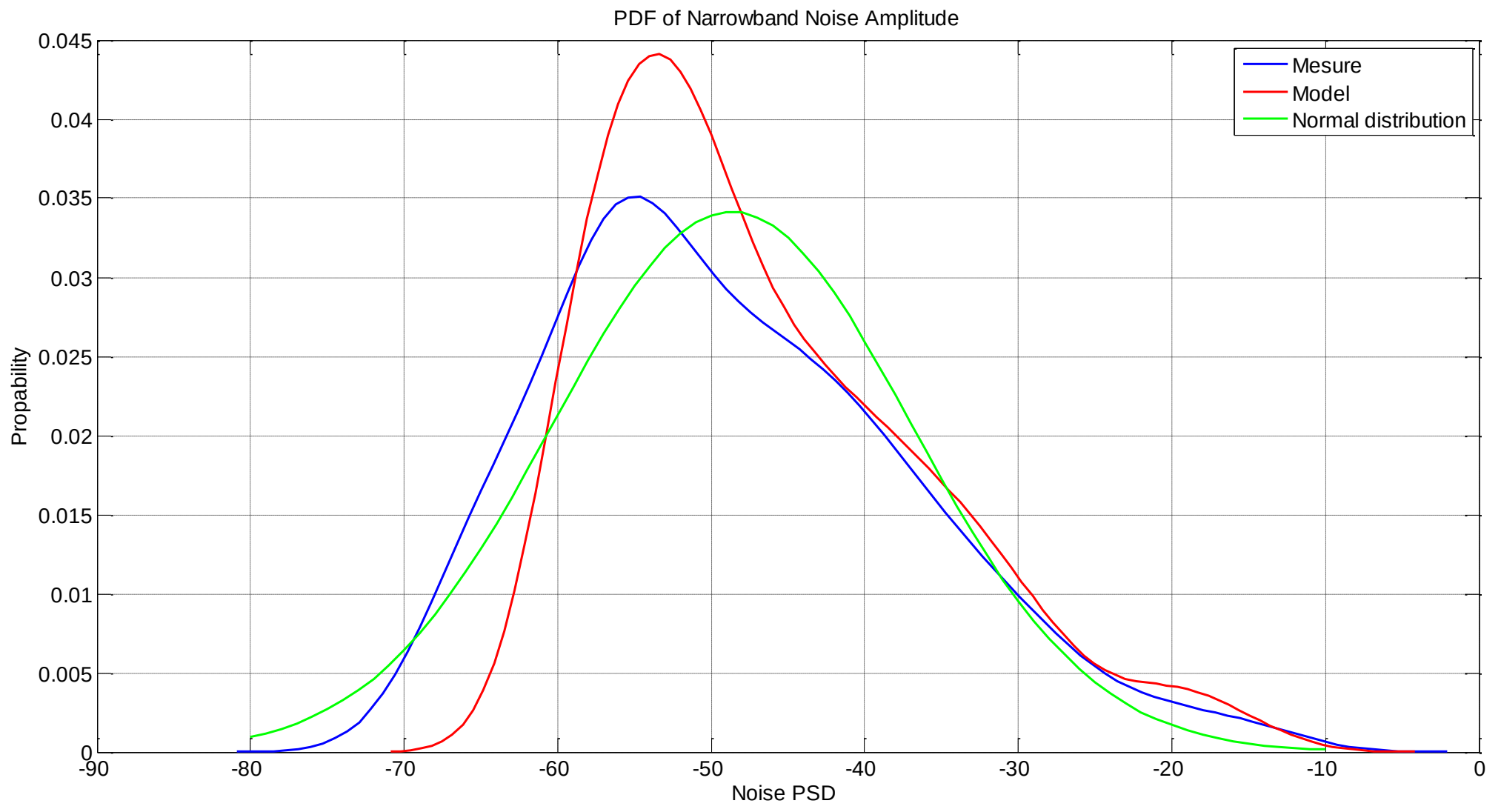

Figure 12. PDF of the modelled and measured noise amplitude.

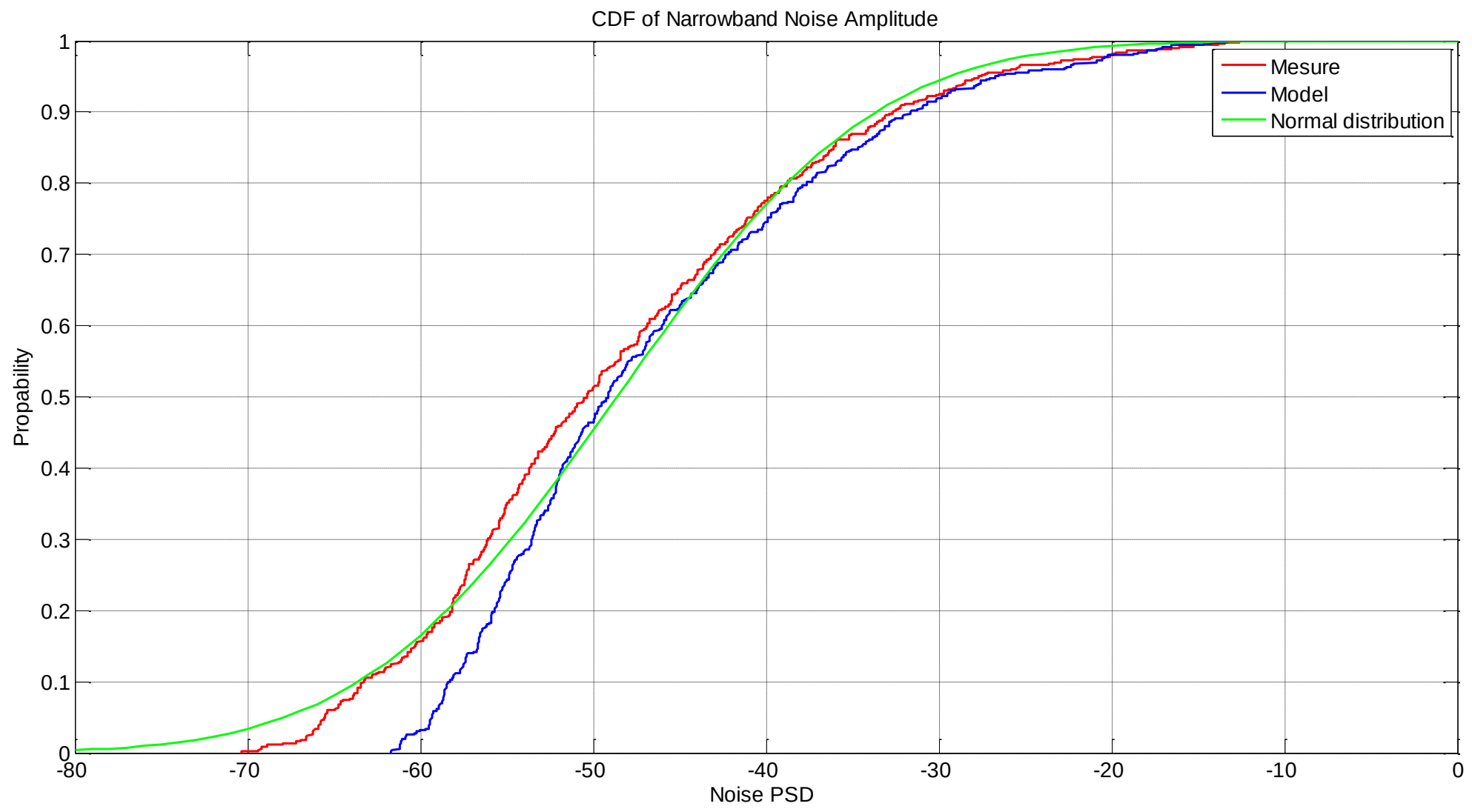

Figure 13. CDF of the modelled and measured noise amplitude.

The comparisons show that the modelled and the measured PDFs are well matched. A similar close match is also achieved for the modelled and measured noise CDF. It is also, observed that simulated noise model has same statistical properties of distribution as the measured noise: In fact, the CDFs of the measured and modelled follow a Gamma distribution w respectively mean value equal to and the Normal distribution fit the PDFs. This proves the relevance of the proposed model.

## 5. Conclusion

In this paper narrowband frequency domain PLC noise is measured and analyzed, and then a statistical characterization and modelling study is presented in detail. Experimental results show that the stationary noise is a decreasing function of frequency with the presence of narrowband interferences. Therefore it considered as a superposition of the colored background noise and narrowband noise. The colored background noise PSD is modelled by a power function, however the narrowband noise PSD is considered as a summation of Gaussian functions. The model of the stationary noise is the sum of these two parametric models. In order to validate the proposed model of the stationary noise a basic statistical analysis is given showing the goodness of the proposed model. In future works, the author will be interested to study the narrowband impulsive noise in low voltage outdoor electrical networks. Then, it become easy to simulate the real PLC system and create a complete OFDM PLC communication system and search for the most appropriate coding and modulation techniques.

## References

[1] A. Haidine, A. Tabone, and J. Muller, " Deployment of power line communication by European utilities in advanced metering infrastructure, "*The 17th International Symposium on Power Line Communications and Its Applications (ISPLC)*, Johannesburg, South Africa, 24-27 March, 2013.

[2] S. Galli, A. Scaglione, and Z. Wang, “Power line communications and the smart grid”, *The first IEEE International Conference on Smart Grid Communications*, Gaithersburg, MD, USA, 4-6 October, 2010.

[3] Dubey, A., et al., “Modeling and Performance Analysis of a PLC System in Presence of Impulsive Noise, ” *IEEE Power & Energy Society General Meeting (PESGM)*, Denver, CO, USA, 26-30 July, 2015.

[4] G. Ndo, F. Labeau and M. Kassouf, “A markov-middleton model for bursty impulsive noise: modeling and receiver design,” *IEEE Transactions on Power Delivery*, vol. 28, no. 4, October 2013.

[5] J.A. Cortes, L. Diez, F.J. Canete and J.J.Sanchez, “Analysis of the indoor broadband power line noise scenario”, *IEEE Transactions on Electromagnetic Compatibility*, vol. 52, no. 4, November 2010.

[6] I.Elfeki, T.Doligez, I. Aouichak, J. Le Bunetel, Y. Raingeaud, “Estimation of PLC transmission line and crosstalk for LV outdoor electrical cables,” in Proceedings of the International Symposium on Electromagnetic Compatibility, Angers, France, September 2017.

[7] C. Kaiser, N. Otterbach, K. Dostert, “Spectral correlation analysis of narrowband power line noise,” In Proceedings of the 2017 IEEE International Symposium on Power Line Communications and its Applications (ISPLC), Madrid, Spain, April 2017

[8] I. Elfeki , S. Jacques, I. Aouichak, T. Doligez, Y.Raingeaud and J. Le Bunetel “Characterization of Narrowband Noise and Channel Capacity for Powerline Communication in France,” Energies 2018.

[9] Mlynek, P., J. Misurec and M. Koutny, “Noise Modeling for Power Line Communication Model, The 35th *International Conference on Telecommunications and Signal Processing (TSP)*, Prague, Czech Republic, 3-4 July 2012.

[10] D’Alessandro, S., M. De Piante and A.M. Tonello, “On Modeling the Sporadic Impulsive Noise Rate within In-Home Power Line Networks, ” *ISPLC*, Austin, TX, USA, 29 March-1 April, 2015.

[11] Hirayama, Y., et al., “Noise Analysis on Wide-band PLC with High Sampling Rate and Long Observation Time,” *ISPLC*. Tokyo, Japan, 2003.

[12] Nyete, A.M., T. Afullo and I.E. Davidson, “Statistical Analysis and Characterization of Low Voltage Power Line Noise for Telecommunication Applications,” *The 12th IEEE AFRICON*, Addis Ababa, Ethiopia, 14-17 September, 2015.

[13] Tiru, B., “Analysis and Modeling of Noise in an Indoor Power Line for Data Communication Purpose,” *Global Research Publications*, January 2012.

[14] Esmailian T. “Multi mega-bit per second data transmission over in-building power lines. Doctoral thesis at University of Toronto. 2003.

[15] OMEGA Deliverable D3.2. PLC Channel Characterization and Modelling. 2008.

[16] H. Philipps, “Performance measurements of powerline channels at high frequencies, ”*The International Symposium on Power Line Communications and Its Applications*, Soka University, Japan, March 1998.

[17] M. Babic, M. Hagenau, K. Dostert, J. Bausch, “Theoretical postulation of PLC channel model,” Open PLC European Research Alliance (OPERA), 2005.

[18] D. Benyoucef, “A new statistical model of the noise power density spectrum for powerline communication,” *International Symposium on Power Line Communications and Its Applications*, Mar. 2003.

[19] R. Alaya and R. Attia, “Coupling Unit for Narrowband Power Line Communications Channel Measurement,”The 24th International Conference on Software, Telecommunications and Computer Networks (SoftCOM